# Comment on "$\mathcal{PT}$-Symmetry Breaking and Laser-Absorber Modes in Optical Scattering Systems"

The Letter by Chong *et al.* [1] presents a theoretical discussion of $\mathcal{PT}$-symmetry of optical systems. In such systems, for two waves propagating in opposite directions under certain conditions, the loss compensation occurs. This happens due to the suppression of the field in dissipating regions and its enhancement in active regions. In Ref. 1, this property is formulated in terms of scattered waves: at particular phase relation between the input waves (eigenvector of the scattering matrix), the sum of the intensities of the waves leaving the system must equal the sum of the intensities of the waves entering the system. The authors stated that this loss compensation occurs for frequencies smaller than the threshold frequency. They show that the system undergoes one or more $\mathcal{PT}$-symmetry breaking transitions as the frequency is tuned up and propose that the effect discussed can be observed in measurements of the spectrum of scattered intensity. In this Comment, we show that the phenomenon discussed in [1] does not occur because $\mathcal{PT}$-symmetry may exist only for isolated frequencies in optical systems.

First, let us note that in any optical system with either loss or gain the frequency dispersion of the permittivity is crucial; in particular, for a gain medium this fact is of the principal importance. Indeed, in a gain medium, if the dispersion is neglected, the possibility of laser generation at high frequencies makes any solution of the linear Maxwell equations incorrect [2]. Laser generation should arise in a sample of any size. Second, due to causality, a dielectric function $\varepsilon(\omega)$ must be analytic in the upper half of the complex frequency plane so that all its singularities are situated in the lower half of the complex plane. Causality must hold for both dissipative and active systems.

Now, let us consider two points **r** and –**r**. We denote the dielectric functions at these points by $\varepsilon_1(\omega)$ and $\varepsilon_2(\omega)$, respectively. $\mathcal{PT}$-symmetry requires that the dielectric functions at points **r** and –**r** be related as $\varepsilon_1(\omega) = \varepsilon_2^*(\omega)$ [1, 3] or

$$\operatorname{Re}\varepsilon_1(\omega) = \operatorname{Re}\varepsilon_2(\omega), \quad \operatorname{Im}\varepsilon_1(\omega) = -\operatorname{Im}\varepsilon_2(\omega). \tag{1}$$

We show that the conditions (1) cannot be satisfied in any finite frequency range. Thus, the threshold $\mathcal{PT}$-symmetry breaking behavior due to the frequency variation (see Fig. 1 in [1]) is not observable.

Let us presume that $\varepsilon_2(\omega)$ is a "causal" dielectric function, so it is analytic in the upper half of the complex plane. After complex conjugation, $\varepsilon_2^*(\omega)$ ceases being analytic and therefore, $\varepsilon_1(\omega)$ cannot be defined as $\varepsilon_2^*(\omega)$ at complex frequency values.

Conditions (1) are formulated for real frequencies for which they are equivalent to complex conjugation. Then, $\varepsilon_1(\omega)$ should be analytically continued to the complex plane. The unique analytic function that satisfies Eqs. (1) is

$$\varepsilon_1(\omega) = \varepsilon_2^*(\omega^*). \tag{2}$$

However, all singularities of $\varepsilon_1(\omega)$ defined by Eq. (2) are in the upper half-plane and the respective response function would, therefore, violate causality.

A similar conclusion can be obtained by using the Kramers-Kronig relations,

$$\operatorname{Re}\varepsilon(\omega) = \varepsilon_0 + \frac{1}{\pi}\mathcal{P}\int_{-\infty}^{+\infty}\frac{\operatorname{Im}\varepsilon(\omega')}{\omega'-\omega}d\omega', \quad \operatorname{Im}\varepsilon(\omega) = -\frac{1}{\pi}\mathcal{P}\int_{-\infty}^{+\infty}\frac{\operatorname{Re}\varepsilon(\omega')-\varepsilon_0}{\omega'-\omega}d\omega'. \tag{3}$$

It is easy to show that if $\varepsilon_2(\omega)$ satisfies Eqs. (3), then $\varepsilon_1(\omega)$ defined by Eqs. (1) does not satisfy Kramers-Kronig relations (3). Thus, a physical system cannot possess properties (1) for an infinite frequency interval. But $\mathcal{PT}$-symmetry for *all* frequencies is not required for the validity of the results of [1]. It would suffice for Eqs. (1) to be valid in a finite frequency range. However, this also is not possible. Indeed, according to the identity theorem, an analytic continuation of $\varepsilon_1(\omega)$ from a finite interval on the real axis to the whole complex half-plane coincides with the analytic continuation of $\varepsilon_1(\omega)$ defined on the whole real axis. Therefore, a dielectric function, which is $\mathcal{PT}$-symmetrical over a finite frequency interval, would not satisfy the Kramers-Kronig relations.

To summarize, $\mathcal{PT}$-symmetry in a finite frequency range violates causality in optical systems. This makes it impossible to observe the $\mathcal{PT}$-symmetry breaking by varying the frequency.

A. A. Zyablovsky,[1] A. P. Vinogradov,[1] A. V. Dorofeenko,[1] A. A. Pukhov,[1] and A. A. Lisyansky[2]

[1]Institute for Theoretical and Applied Electromagnetics RAS, 13 Izhorskaya, Moscow 125412, Russia
[2]Department of Physics, Queens College of the City University of New York, Flushing, NY 11367, USA